\documentclass[conference]{IEEEtran}
\IEEEoverridecommandlockouts
\usepackage[numbers,sort&compress]{natbib}
\usepackage{amsmath,amssymb,amsfonts}
\usepackage{algorithmic}
\usepackage{graphicx}
\usepackage{xcolor}
\usepackage{multirow}
\usepackage{makecell}
\usepackage{stfloats}
\usepackage{amsmath}
\usepackage{textcomp}
\usepackage[T1]{fontenc}

\usepackage{url}
\def\BibTeX{{\rm B\kern-.05em{\sc i\kern-.025em b}\kern-.08em    T\kern-.1667em\lower.7ex\hbox{E}\kern-.125emX}}
\begin{document}

\title{A Model for Assessing Network Asset Vulnerability Using QPSO-LightGBM\
}

\author{\IEEEauthorblockN{Xinyu Li}
\IEEEauthorblockA{\textit{$S^2$AC Lab,School of Computer and Information} \\
\textit{Hefei University of Technology}\\
Hefei,China \\
XinYuLi@mail.hfut.edu.cn}
\and
\IEEEauthorblockN{Yu Gu}
\IEEEauthorblockA{\textit{$S^2$AC Lab,School of Computer and Information} \\
\textit{Hefei University of Technology}\\
Hefei,China \\
yugu.bruce@ieee.org}
\and
\IEEEauthorblockN{Chenwei Wang}
\IEEEauthorblockA{\textit{$S^2$AC Lab,School of Computer and Information} \\
\textit{Hefei University of Technology}\\
Hefei,China \\
ChenWeiWang@mail.hfut.edu.cn}
\and
\IEEEauthorblockN{Peng Zhao}
\IEEEauthorblockA{\textit{$S^2$AC Lab,School of Computer and Information} \\
\textit{Hefei University of Technology}\\
Hefei,China \\
2017212168@mail.hfut.edu.cn}
}

\maketitle

\begin{abstract}
With the continuous development of computer technology and network technology, the scale of the network continues to expand, the network space tends to be complex, and the application of computers and networks has been deeply into politics, the military, finance, electricity, and other important fields. When security events do not occur, the vulnerability assessment of these high-risk network assets can be actively carried out to prepare for rainy days, to effectively reduce the loss caused by security events. Therefore, this paper proposes a multi-classification prediction model of network asset vulnerability based on quantum particle swarm algorithm-Lightweight Gradient Elevator (QPSO-LightGBM). In this model, based on using the Synthetic minority oversampling technique (SMOTE) to balance the data, quantum particle swarm optimization (QPSO) was used for automatic parameter optimization, and LightGBM was used for modeling. Realize multi-classification prediction of network asset vulnerability. To verify the rationality of the model, the proposed model is compared with the model constructed by other algorithms. The results show that the proposed model is better in various predictive performance indexes.
\end{abstract}


\section{Introduction}

New technologies such as big data, cloud computing, the Internet of Things, artificial intelligence, and blockchain continue to emerge, and human society is accelerating into the era of the digital economy. Cyber security incidents keep cropping up, organized, targeted forms of cyber attacks are becoming ever more apparent, the network security risk continues to increase, and the network security is not just about national security, social security, urban security, infrastructure, security, and also closely related to everyone's life, Cyberspace is regarded as the fifth frontier after land, sea, air, and sky, and has become a new battlefield for countries to play games and its importance is comparable to that of oil resources\cite{alperovitch2011towards}. Including the United States, Russia, Britain more countries around the world will be raised to a new height, network security in our country is also increasingly paying attention to cyber security, March 11, 2021, 13th session of the National People's Congress four conference by the law of the People's Republic of China on the national economic and social development of 14 five-year plan and 2035 vision outline, Which referred to the "network security" 14 times, "data security" 4 times, involving the digital economy, digital ecology, national security, energy and resources security four fields, the network security industry has been determined in the national policy level will be further "nurtured and strengthened".

Cyberspace mapping technology is an effective means to identify and control the elements of cyberspace, prevent network threats and vulnerabilities, and improve network security, which is of great value to the construction of a network security system. The United States is the first country to promote cyberspace mapping applications and has now formed a complete cyberspace detection infrastructure and system.Representative projects includes the US Defense Agency's X program\cite{nakashima2012plan}, the US Bureau of Lands' SHINE program\cite{grant2015military}, the US National Security Agency's Treasure Map\cite{rashid2006extending}, and Censys, Shodan \cite{matherly2015complete}and etc. China has also made some deployment and research on the direction of cyberspace surveying and mapping. Representative examples include ZoomEye of Know Chuangyu, FOEYE of Huashun Xinan\cite{zhang2016design}, and ``Network Asset Mapping and Analysis System"\cite{kou2022research} and etc.

With the continuous development of network space assets detection technology, cyberspace assets-related data is exposed to the public, which to a certain extent, is an opportunity for criminals, of cyberspace assets security to pose a threat. To better maintain cyberspace assets, this paper carries out a vulnerability assessment on cyberspace assets to find vulnerable assets\cite{crampton2003political}. Asset vulnerability refers to the defects and deficiencies of network assets exposed to the Internet, which will indirectly or directly cause harm to cyberspace assets. Based on the vulnerability assessment of network assets, this paper prioritizes the vulnerability of network assets, to minimize the probability of security events \cite{10149418} and reduce the loss caused by the occurrence of security events.

The existing methods of vulnerability assessment are to score vulnerabilities. The most representative vulnerability assessment method is the CVSS Universal vulnerability scoring system \cite{elbaz2020fighting}. In 2008, 5 SCADA industry experts proposed a cyber terrorism SCADA risk assessment framework system \cite{beggs2008proposed}. In 2020, Kitty Kioskli et al. achieved a qualitative assessment of network risk\cite{kioskli2020socio}. In 2020, Hua Dong et al. proposed a smart grid information security Risk assessment (ISRA) method\cite{dong2020combination}. In 2021, Maček D et al. proposed a hybrid multi-criteria model for critical IT system evaluation with risk analysis and evaluation elements as evaluation criteria\cite{mavcek2021model}.

The above research mainly quantifies or qualifies the properties of cyberspace assets to assess the vulnerability of assets. But in the field of surveying and mapping in cyberspace, there is no space mapping system based on network vulnerability assessment of relevant research \cite{9613773}, therefore, this paper is based on the detection results of the existing network space mapping system, through the analysis of the vulnerability of assets related properties, combined with LightGBM decision tree algorithm, based on the balance of data processing, Assess the vulnerability of cyberspace assets.

\section{Related works}
\subsection{Research Status of Cyberspace Mapping Technology}
Cyberspace mapping technology is an effective means to identify and control the elements of cyberspace, prevent network threats and vulnerabilities, and improve network security, which is of great value to the construction of a network security system \cite{9484767}. This technology is an innovative technology for real description and intuitive reflection of cyberspace. It is a cyclic process of detection, collection, processing, analysis, and application. Based on computer science, surveying and mapping science, network science, and information science, it takes cyberspace assets as the research object \cite{10339891}. Measurement, physical location, geography, surveying and mapping through the network and other related information visualization theory and means of science and technology, access to the network attribute space assets of cyberspace and the geographical space, the network information such as the topology and the environment of space assets through the visual form, to build a comprehensive global Internet network space mapping system.

The United States is the first country to conduct cyberspace surveying and mapping and has formed a relatively complete cyberspace exploration system. The most prominent examples are the SHINE program of the Ministry of Land and Resources (DHS), the TreasureMap program of the National Security Agency (NSA), and the X Program of the Department of Defense's Advanced Research Projects Agency (DRAPA). China has also made some deployment and research on the direction of cyberspace surveying and mapping, under the support of key R\&D projects and other projects of the national Ministry of Science and Technology, and has gained rich research results. In terms of system design, representative examples include the cyberspace mapping system of China Electronic Science and Technology Network Information Security Co., LTD., ZoomEye of Know Chuangyu, FOEYE of Huashun Xinan, and "Network Asset Mapping and Analysis System" designed and developed by the First Research Institute of the Ministry of Public Security, etc.

Cyberspace mapping technology provides abundant data related to cyberspace assets. On this basis, we can assess the vulnerability of cyberspace assets more smoothly.

\subsection{Research Status of Vulnerability Assessment}
Information systems or assets exposed to the Internet are more likely to be exploited by lawbreakers due to their vulnerabilities and vulnerabilities. Vulnerability assessment is mainly a process of the comprehensive inspection of these systems or assets, screening out the systems or assets with security problems, and ranking the possibility of these security problems being used by criminals. Through the critical infrastructure industry information system or asset vulnerability assessment, their level of security risks can be analyzed and sorted according to the results of the grade evaluation, before security vulnerability rating higher asset maintenance, thus effectively reducing the possibility of security problems and losses, better protection information system or asset.

The existing methods of vulnerability assessment are to score vulnerabilities. The most representative vulnerability assessment method is the CVSS Universal vulnerability scoring system \cite{elbaz2020fighting}. In 2008, 5 SCADA industry experts proposed a cyber terrorism SCADA risk assessment framework system \cite{beggs2008proposed}. In 2020, Kitty Kioskli et al. achieved a qualitative assessment of network risk\cite{kioskli2020socio, 10443215}. In 2020, Hua Dong et al. proposed a smart grid information security Risk assessment (ISRA) method\cite{dong2020combination}. In 2021, Maček D et al. proposed a hybrid multi-criteria model for critical IT system evaluation with risk analysis and evaluation elements as evaluation criteria\cite{mavcek2021model, 10251628}.

The above research mainly quantifies or qualifies the properties of cyberspace assets to assess the vulnerability of assets. But in the field of surveying and mapping in cyberspace, there is no space mapping system based on network vulnerability assessment of relevant research, therefore, this paper is based on the detection results of the existing network space mapping system, through the analysis of the vulnerability of assets related properties, combined with LightGBM decision tree algorithm, based on the balance of data processing, Assess the vulnerability of cyberspace assets. GBDT (Gradient Boosting Decision Tree) is a classical model in machine learning. The main idea of GBDT is to use a weak classifier (Decision Tree) to iteratively train to get the optimal model, which has the advantages of a good training effect and is not easy to overfit\cite{chen2015xgboost}. LightGBM(Light Gradient Boosting Machine) is a framework to implement the GBDT algorithm. LightGBM's current studies on evaluation include network warfare simulation and effectiveness evaluation, credit rating evaluation, real estate price evaluation and etc. The algorithm supports efficient parallel training and has the advantages of faster training speed, lower memory consumption, better accuracy, and fast processing of massive data. Based on the above characteristics, this paper intends to use the LightGBM model to evaluate the vulnerability of cyberspace assets\cite{ke2017lightgbm}.

\section{Data and Processing Methods}
\subsection{Source Data}
\subsubsection{Data Collection and Processing}
In this paper, the web crawler technology is adopted to obtain the data of cyberspace assets in each system by calling the open interfaces in cyberspace mapping systems such as Censys, Shodan, Fofa, and 360Quake. By crawling the data of the four platforms, the obtained IP number and port number are used as the unique ID number of the cyberspace asset to identify it. After removing the data with seriously missing attributes, this experiment finally collected 24,000 network asset data with relatively complete attributes, and each data contained 110 attributes, including IP, port number, domain name, industry, region, vulnerability type, certificate, etc.

Due to the properties of network assets collected data contains too much, among them, the part between the properties and assets of fragility, there is no direct relationship, if the asset directly all of the attributes of the feature of vulnerability evaluation experiment and training can lead to a large amount of calculation, time is too long, data redundancy, and due to the influence of many irrelevant attributes, will seriously affect the predicted results of the model. Therefore, before data training, it is necessary to screen the attributes of assets first, and extract the attributes related to the vulnerability of assets as data for subsequent experiments, to reduce the workload of training and improve the training effect of the model. The selection of features in this paper mainly refers to the Implementation Guide for Telecommunications Network and Internet Data Security Risk Assessment, and the characteristics of network assets are divided into three parts according to the characteristics of network assets data: management factors, technical factors and vulnerability factors, and the vulnerability characteristics of network assets are extracted, as shown in Table~\ref{tab1}:

\begin{table*}[tp]
\caption{Vulnerability Related Attributes}
\begin{center}
\begin{tabular}{|c|c|c|c|}
\hline
\textbf{Classification} & \textbf{Characteristics}& \textbf{Instructions}& \textbf{The Instance} \\
\hline
\multirow{4}*{\makecell[c]{Management\\factors} }& Weak password & Whether a weak password exists & NO \\
\cline{2-4} 
& Firewall & Whether network assets are protected by firewalls & NO \\
\cline{2-4} 
& Cloud hosting & Whether to set up on a cloud host & NO \\
\cline{2-4} 
& CDN & Whether the network asset has the CDN technology& NO \\
\cline{1-4} 
\multirow{5}*{\makecell[c]{Technical \\factors}}& Operating System Model & computer program model & Ubuntu18.04 \\
\cline{2-4} 
& Website development language model & A development language for websites on Web properties & PHP5.3.29 \\
\cline{2-4} 
& Web Container Model &The model of the server& Apache2.4.33 \\
\cline{2-4} 
& The number of fingerprints detected & The number of fingerprints of an asset& 4 \\
\cline{2-4} 
& Web Application Model &The model of the application& WordPress5.9.3 \\
\cline{1-4} 
\multirow{3}*{\makecell[c]{Vulnerability \\factors}}& CVSS scores & CVSS score for POC-validated 
vulnerabilities on the asset & 10 \\
\cline{2-4} 
& Number of existing vulnerabilities & number on the network asset &CVE-2021-44228 \\

& Vulnerability discovery time & Poc-verified vulnerability discovery time on network assets & 2022/04/10 \\
\hline
\end{tabular}
\label{tab1}
\end{center}
\end{table*}

As shown in Table~\ref{tab1}, the characteristics of the data set contain some numeric data features and more character data, it will not be easy to use the LightGBM model of data for training, to decrease the complexity of the subsequent model training and improve the accuracy of the training results, need before training model to quantize the character data type, The specific processing, and coding methods As shown in Table~\ref{tab2}. Since there are many categories in the features, the common one-hot Encoding method may cause dimensional disaster. Therefore, Label Encoding is used to numerically process the features in this paper. In addition, due to the operating system, web development language, the web container, and web application the four characteristics of categories are overmuch, direct use of the Label-Encoding Encoding method will lead to a large array index, the data conversion cost is too high, so we need to the concept of these attributes, namely on the premise of least affected the result of the vulnerability assessment, The existing categories of data are merged appropriately to reduce the number of feature categories and facilitate feature coding.

\begin{table*}[tp]
\caption{Feature Coding and Processing Methods}
\begin{center}
\begin{tabular}{|c|c|c|c|}
\hline
\textbf{Characteristics to be addressed} & \textbf{process mode}& \textbf{The sample}& \textbf{encoding} \\
\hline
Weak password & NO & NO & Label Encoding \\
\hline
firewall & NO & NO & Label Encoding \\
\hline
Cloud hosting & NO & NO & Label Encoding \\
\hline
CDN & NO & NO & Label Encoding \\
\hline
The operating system & concept & Ubuntu18.04→Ubuntu18 & Label Encoding \\
\hline
Web Development language & concept & PHP5.3.29→PHP5.3 & Label Encoding \\
\hline
The web container & concept & Apache2.4.33→Apache2.4 & Label Encoding \\
\hline
The web application & concept & WordPress5.9.3→WordPress5.9 & Label Encoding \\
\hline
Number of existing vulnerabilities & concept & CVE-2021-44228→CVE-2021 & Label Encoding \\
\hline
Vulnerability discovery time & Convert to a value based on 1900/1/1 & 2022/4/7→44658 & NO \\
\hline
\end{tabular}
\label{tab2}
\end{center}
\end{table*}

\subsubsection{Data Vulnerability Labels}
Since there is no vulnerability score for the collected network asset data, this paper selects the expert scoring method to mark the vulnerability of the data. The marking range is 0-10, where 0 to 10 indicates the vulnerability from weak to strong, and the training dataset of the experiment in this paper is obtained after the expert scoring. The steps of expert scoring are composed of four parts: expert selection, expert scoring, comprehensive expert score, and expert review. In the expert selection stage, five representative and authoritative experts in the field of cyberspace security are selected. The selected experts must be familiar with and master the assessment criteria and process of asset vulnerability. In the expert scoring stage, each expert evaluates the vulnerability of assets independently, and the evaluation results are summarized after all the experts have completed the evaluation. In the comprehensive expert score stage, the scores of each expert should be comprehensively scored according to the voting method, and the final score closest to the vulnerability of the asset itself can be obtained. In the expert review stage, experts are summoned to review the vulnerability of the asset. If there is no objection from the experts, the vulnerability score of the asset will be passed; otherwise, the final vulnerability value of the asset will be discussed by experts according to the rule of minority subordination. The final result of data input for model training is shown in Table~\ref{tab3}.
\begin{table*}[tp]
\caption{Vulnerability Dataset of Network Assets}
\begin{center}
\begin{tabular}{|c|c|c|c|c|c|c|c|c|}
\hline
\textbf{OS} & \textbf{Web\_container}& \textbf{Web\_app}& \textbf{num\_assembly}& \textbf{...}& \textbf{firewall}& \textbf{C\_hosting} & \textbf{CDN} & \textbf{Score} \\
\hline
8 & 0 & 12 & 3 & ... & 0 & 1 & 0 & 2 \\
\hline
19 & 0 & 2 & 2 & ... & 0 & 1 & 0 & 4 \\
\hline
9 & 0 & 16 & 0 & ... & 0 & 0 & 1 & 5 \\
\hline
11 & 0 & 19 & 0 & ... & 0 & 1 & 0 & 0 \\
\hline
6 & 2 & 25 & 3 & ... & 0 & 1 & 0 & 4 \\
\hline
... & ... & ... & ... & ... & ... & ... & ... & ... \\
\hline
3 & 0 & 9 & 4 & ... & 1 & 1 & 0 & 10 \\
\hline
\end{tabular}
\label{tab3}
\end{center}
\end{table*}
\subsection{Data Imbalance Processing}

\subsubsection{Data Imbalance Processing}
\begin{figure}[hp]
\centerline{\includegraphics[width=0.15\textwidth,height=0.15\textwidth]{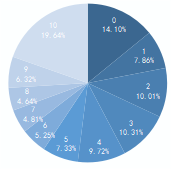}}
\caption{Vulnerability Value Distribution before Treatment.}\label{fig:1}
\label{fig1}
\end{figure}

Data imbalance, also known as "data skew", mainly refers to the significant differences in the distribution of samples of different categories, which may lead to the performance degradation of learning algorithms\cite{li2019review}. Therefore, before model training, if the data distribution is unbalanced, it is necessary to deal with the imbalanced training data first, to maximize the training effect of the model. In this paper, the vulnerability value of network assets is divided into 11 levels from 0 to 10, and the experimental data sets are divided into 11 categories according to the 11 levels. The proportion of each category of vulnerability in the experimental data sets is shown in Figure~\ref{fig1}.

\begin{figure}[h]
\centerline{\includegraphics[width=0.15\textwidth,height=0.15\textwidth]{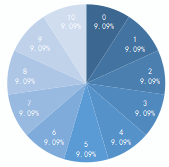}}
\caption{Distribution of Vulnerability Values after Treatment.}\label{fig:2}
\label{fig2}
\end{figure}

As can be seen from Figure~\ref{fig1}., some categories of data are nearly \(20\%\), and some categories of data are only less than \(5\%\). If such unbalanced data sets are directly evaluated, underfitting of a small number of samples may occur, while overfitting of a large number of samples may affect the accuracy of the model to a certain extent. SMOTE oversampling is a common method to deal with unbalanced data. Therefore, this paper uses SMOTE oversampling to deal with data. The vulnerability value distribution after processing is shown in Figure~\ref{fig2}.

\section{Network Asset Vulnerability Assessment Model Based on QPSO-LightGBM}
\subsection{Introduction to LightGBM Algorithm}
The main idea of the Gradient Boosting Decision Tree (GBDT)\cite{chen2016xgboost} is to use iterative training of the Decision Tree to obtain the optimal model. LightGBM\cite{ke2017lightgbm} is a framework to implement the GBDT algorithm. Based on Extreme Gradient Boosting (XGBoost)\cite{chen2015xgboost}, It has optimized the decision tree algorithm based on histogram, Leaf growth strategy, Cache hit ratio optimization, and sparse feature multi-thread optimization based on the histogram and etc, which makes it have better computational performance, less memory consumption, and better overall efficiency. The function is shown in the equation(1) \cite{wu2019improved}, where \(R_i\) is the true value of labels, \(C^{k-1}\) is the sum of the regularization terms of the first k-1 trees, and \(\hat{R}_i^k\) is the result of the KTH learning. The meaning of the objective function is to find a tree \(T_k\) that minimizes the value of the function.

\begin{equation}
\makecell[c]{Object^{k} = \sum\limits_{i}L(R_{i} + \hat{R}_{i}^{k}) + \omega(T_{k}) + C^{k-1} \\ \\= \sum\limits_{i}L(R_{i} , \hat{R}_{i}^{K} + T_{k}(x_i)) + \omega(T_k) + C^{k-1}}\label{eq:1}
\end{equation}

Taylor's formula is used to expand the objective function:

\begin{equation}
T(x + \vartriangle{x}) = T(x) + T^{'}(x)\vartriangle{x} + \frac{1}{2}T^{''}(x)(\vartriangle{x})^2\label{eq:2}
\end{equation}

The result of second-order Taylor formula expansion of the loss function is:

\begin{equation}
\makecell[c]{\sum\limits_{i}L(R_{i} , \hat{R}_{i}^{K} + T_{k}(x_i)) = \sum\limits_{i}[L(R_{i} + \hat{R}_{i}^{k-1}) \\ \\+ L^{'}(R_{i} + \hat{R}_{i}^{k-1})]T_{k}(x_i) + \frac{1}{2}L^{''}(R_{i},\hat{R}_{i}^{k-1}T_{k}(x_i))}\label{eq:3}
\end{equation}

Write \(g_i\) the first derivative of the ith sample loss function, and \(h_i\) the second derivative of the ith sample loss function:

\begin{equation}
\makecell[c]{g_{i} = L^{'}(R_{i} , \hat{R}_{i}^{k-1})}\label{eq:4}
\end{equation}

\begin{equation}
\makecell[c]{h_{i} = L^{''}(R_{i} , \hat{R}_{i}^{k-1})}\label{eq:5}
\end{equation}
Then the objective function can be simplified as:

\begin{equation}
\makecell[c]{Object^{k} = \sum\limits_{i}[L(R_{i} , \hat{R}_{i}^{k-1}) + g_{i}T_{k}(x_{i}) + \frac{1}{2}h_{i}T_{k}^{2}(x_{i})] \\ \\+ \omega(T_{k}) + c}\label{eq:6}
\end{equation} 

\subsection{Quantum Particle Swarm Optimization}
Particle swarm optimization (PSO) is an optimization algorithm based on group cooperation, which is widely used in parameter optimization due to its advantages of fast convergence speed and high optimization accuracy\cite{rokbani2021beta}. In the standard PSO algorithm system, the velocity and position of the particle at a certain moment are related to the velocity and position of the particle at the previous moment, which determines that the velocity and position of the particle at any time do not have randomness, resulting in the search area of the particle can not cover all feasible space, and it is easy to fall into local extremum. Quantum Particle Swarm Optimization (QPSO) is a new type of PSO algorithm based on DELTA potential. In this algorithm, particles do not have the property of moving direction, the particles follow the random rules of quantum motion, and the current particle motion state is not affected by the previous time. Therefore, compared with the standard PSO algorithm, particles with quantum behavior have better global search performance and can better converge to the global optimal solution.

The velocity and position of a particle with quantum behavior are uncertain, but its motion state can be represented by the probability density function of the particle appearing at a certain point in the search space, which can be replaced by the square of the wave function. The probability density can be obtained by solving the Schrodinger equation, and the exact position of the particle can be obtained by Monte Carlo simulation. The position equation is as follows:

\begin{equation}
X(t+1) = P \pm \beta|P_{mbest} - X(t)|\ln(\frac{1}{\mu})\label{eq:7}
\end{equation}

Where: P =(P1, P2... PN) is the random point where the particle moves in the feasible space; T is the current iteration number; X(t) is the position vector of the current particle at time t; $\beta$ is the shrinkage-expansion coefficient; $\mu$ is a random number. \(P_{mbest}\) is the average optimal position of global particles, and its calculation formula is as follows:

\begin{equation}
\makecell[c]{P_{mbest} = \frac{1}{M} = \sum\limits_{i=1}^{M}P_{i} = \\ \\(\frac{1}{M}\sum\limits_{i=1}^{M}P_{i,1},\frac{1}{M}\sum\limits_{i=1}^{M}P_{i,2},...,\frac{1}{M}\sum\limits_{i=1}^{M}P_{i,N})}\label{eq:8}
\end{equation}

Where: M is the total number of particles in the particle swarm in the feasible space; N is particle dimension; \(P_i\) is the individual optimal position of the ith particle. In the process of particle movement,satisfying the following formula:

\begin{equation}
P_i(t) = \left\{
	\begin{aligned}
	X_{i}(t) && \emph{f}[X_{i}(t) < \emph{f}P_{i}(t-1)]\\
	P_{i}(t-1) && \emph{f}[X_{i}(t) \geq \emph{f}P_{i}(t-1)]\\
	\end{aligned}
	\right.\label{eq:9}
\end{equation}

Where,\(\emph{f}[X_{i}(t)]\) is the fitness function. Thus, the global optimal position of PSO can be determined, that is,the expression of searching the optimal solution is:

\begin{equation}
P_{g}(t) = \mathop{argmin}\limits_{1\le i \le m}{\emph{f}[P_{i}(t)]}\label{eq:10}
\end{equation}

The formula shows that the particle state in QPSO is only represented by the position vector X(t), and only one parameter needs to be adjusted during the execution of the algorithm. Therefore, QPSO has a faster convergence speed compared with standard PSO.

\subsection{QPSO-LightGBM Algorithm}
In this paper, the LightGBM algorithm is used to implement multi-classification tasks for datasets. Due to the large number of parameters of the algorithm, and some parameters have a certain influence on the evaluation results, the QPSO algorithm has unique advantages in optimizing the parameters of LightGBM, which can effectively improve the evaluation effect of the model. Therefore, in this experiment, several parameters affecting the high accuracy of the LightGBM multi-classification evaluation model were optimized by the QPSO algorithm in this paper. The information on each parameter and the optimization range are shown in Table~\ref{tab4}.

\begin{table*}[bp]
\caption{Parameters Information And Range To Be Optimized}
\begin{center}
\begin{tabular}{|c|c|c|}
\hline
\textbf{parameters} & \textbf{Parameters of the content}& \textbf{The parameter range} \\
\hline
learning rate & Model training learning rate & [0.01, 0.2] \\
\hline
n estimators & Number of model training iterations & [1000, 3000] \\
\hline
max depth & Maximum depth of tree model & [5, 12] \\
\hline
num leaves & The number of leaves on a tree & (1, 1024) \\
\hline
feature fraction & Create the feature sampling ratio for the tree tree & [0.5, 1.0] \\
\hline
bagging fraction & Create the data sampling ratio for the tree & [0.5, 1.0] \\
\hline
\end{tabular}
\label{tab4}
\end{center}
\end{table*}

\subsection{Dismantling Method}
\begin{figure}[h]
\centerline{\includegraphics[width=0.4\textwidth,height=0.2\textwidth]{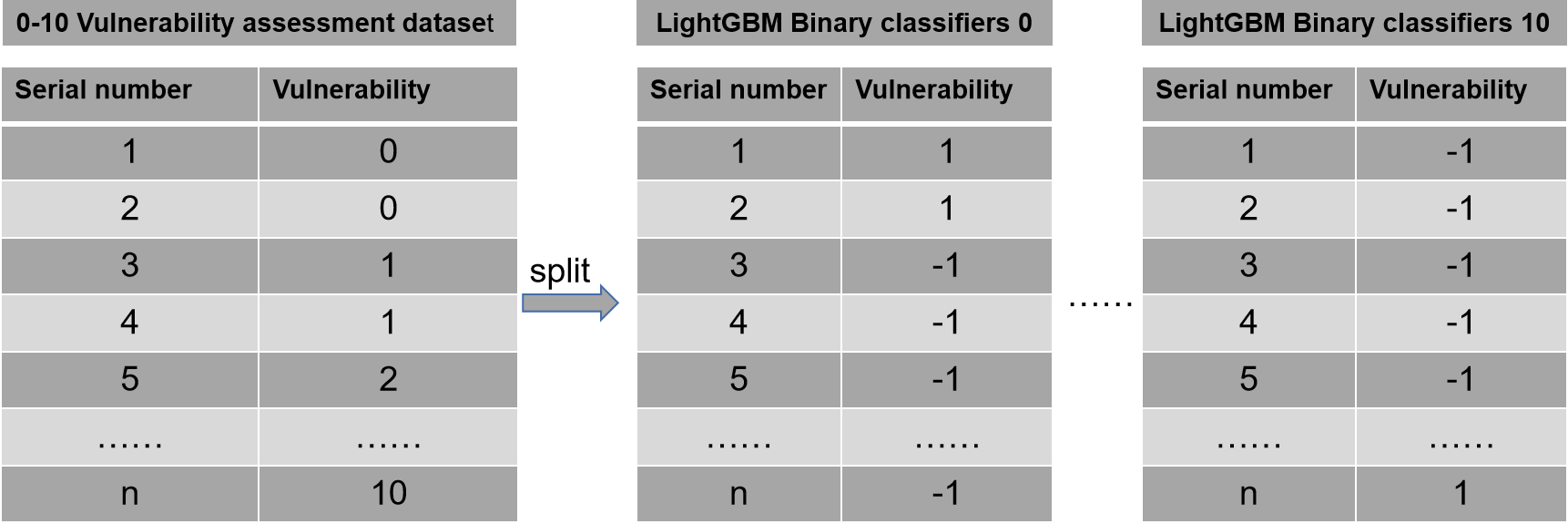}}
\caption{Division of disassembly method}\label{fig:3}
\end{figure}

As LightGBM is a decision tree algorithm, the accuracy of this algorithm in realizing multiple classifications is far lower than that of realizing binary classification. Therefore, this paper adopts the disassembly method to achieve the final multi-classification model prediction by combining multiple binary classification algorithms, to improve the accuracy of model classification. In this paper, the vulnerability assessment range is 0-10, a total of 11 categories. Therefore, we need to build 11 LightGBM classifiers, and the training task of each classifier is shown in Figure~\ref{fig:3}. The vulnerability assessment model constructed by 11 LightGBM binary classifiers generates 11 training results-1 or 1 for each type of vulnerability value during training, and the vulnerability value is encoded by these 11 values.

\section{Experimental Results And Analysis}
\subsection{Evaluation Index}
There are mainly four performance indicators used to evaluate the classification model: Accuracy, Precision, Recall, and F1 score. In this paper, the accuracy rate represents the percentage of correct results in the total number of data sets for assessing the vulnerability of network assets. The accuracy rate is the probability that the asset of each vulnerability value is the predicted value in the model prediction result. Represents the prediction accuracy of the model; The calculation formula of accuracy and precision rate is shown in the equation(11)-equation(12). In the definition of accuracy and precision rate, TP positive sample is judged as positive, TN negative sample is judged as negative, FP negative sample is judged as positive, FN positive sample is judged as negative.
\begin{equation}
Accuracy = \frac{TP + TN}{TP + TN + FP + FN}\label{eq:11}
\end{equation}
\begin{equation}
Precision = \frac{TP}{TP + FP}\label{eq:12}
\end{equation}

Since the vulnerability value distribution of the experimental data set in this paper is unbalanced, the model cannot be evaluated only by referring to the accuracy rate. Recall and F1 scores should be introduced to evaluate the model, and the calculation formula is shown in the \eqref{eq:13}--\eqref{eq:14}. The recall ratio refers to the proportion of assets with correct vulnerability assessment among all assets assessed as a certain category. The F1 score is the harmonic average of precision and recall.
\begin{equation}
Recall = \frac{TP}{TP + FN}\label{eq:13}
\end{equation}
\begin{equation}
\frac{2}{F_{1}} = \frac{1}{Precision} + \frac{1}{Recall}\label{eq:14}
\end{equation}

However, the experimental vulnerability value in this paper has been classified 11 times. so in this paper, according to the Macro business rules to calculate, the predicted results computed each time the sort of precision ratio and recall ratio and F1 score, finally, take the mean value, so that each classification evaluation is treated equally. Thus, the precision rate, recall rate, and F1 score are shown in the \eqref{eq:15}--\eqref{eq:17}, where n is the total number of types of vulnerability value, namely 11.
\begin{equation}
Precision_{Macro} = \frac{1}{n}\sum_{i=1}^{n}Precision_{i}\label{eq:15}
\end{equation}
\begin{equation}
Recall_{Macro} = \frac{1}{n}\sum_{i=1}^{n}Recall_{i}\label{eq:16}
\end{equation}
\begin{equation}
F_{Macro} = \frac{1}{n}F_{i} = \frac{2 \times Precision_{Macro} \times Recall_{Macro}}{Recall_{Macro} + Precision_{Macro}}\label{eq:17}
\end{equation}

\subsection{Results and Discussion}
In this paper, the acquired network asset data is processed above to form a data set containing 24000 samples and 12 characteristic variabe The category labels are divided into 11 grades from 0 to 10 based on the expert scoring results. Select \(80\%\) of the data as the training set and \(20\%\) as the test set to make a multi-classification prediction of the vulnerability of network assets. The experimental results are shown in Table~\ref{tab5}, which shows the evaluation index values of the classification of network asset vulnerability using the QPSO-LightGBM model.

\begin{table}[hp]
\caption{The evaluation indicators value of each model}
\begin{center}
\begin{tabular}{|c|c|c|c|}
\hline
\textbf{accuracy} & \textbf{\makecell{accurate rate}}& \textbf{\makecell{The recall  rate}}& \textbf{\makecell{F1 Score}} \\			
\hline
93.19\(\%\) & 93.25\(\%\) & 93.19\(\%\) & 93.18\(\%\) \\
\hline
\end{tabular}
\label{tab5}
\end{center}
\end{table}

\begin{figure}[hp]
\includegraphics[width=7.5 cm]{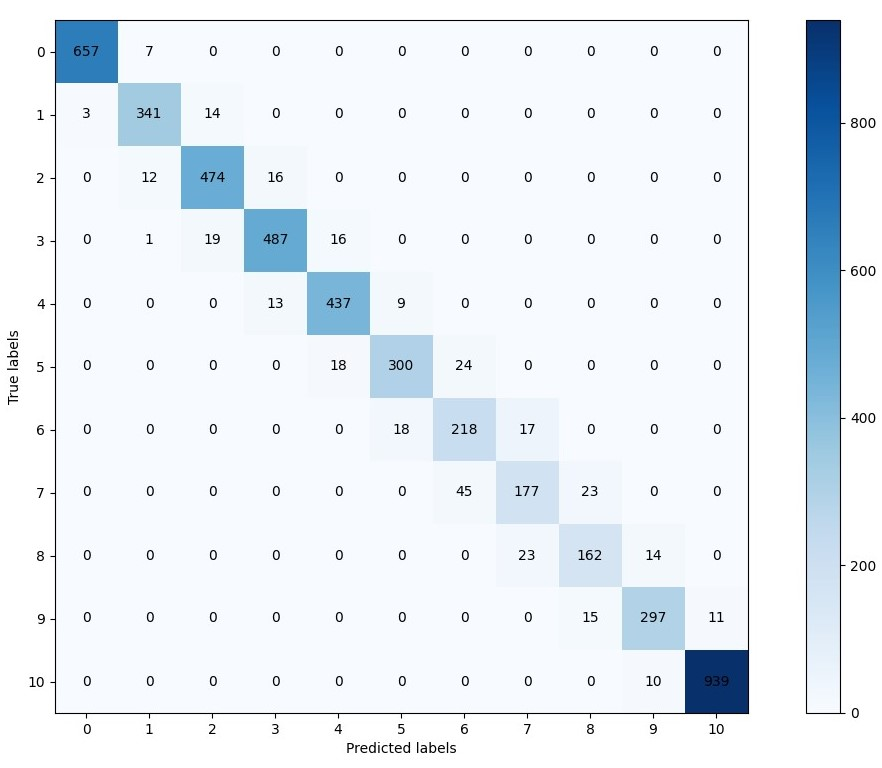}
\caption{Confusion Matrix Of QPSO-LightGBM Model.\label{fig4}}
\end{figure}

The confusion matrix of the prediction results was visualized, and the confusion matrix of the QPSO-LightGBM model was obtained as shown in Figure~\ref{fig4}. It can be seen from Table~\ref{tab5} and Figure~\ref{fig4} that the QPSO-LightGBM model performs well in all evaluation indexes on the network asset vulnerability dataset

\unskip

\subsection{Comparison of Different Algorithms}
In order to further verify the superiority of QPSO-LighTGBM model in the performance of network asset vulnerability classification, this paper conducts a comparative study on this model with LightGBM, XGBoost, GBDT, SVM and Random Forest. The comparative experimental results of each model are shown in Table~\ref{tab6}.

\begin{table}[hp]
\caption{The evaluation indicators value of each model}
\begin{center}
\begin{tabular}{|c|c|c|c|c|}
\hline
\textbf{model} & \textbf{accuracy} & \textbf{\makecell[c]{accurate \\rate}}& \textbf{\makecell[c]{The recall\\ rate}}& \textbf{\makecell[c]{F1 \\Score}} \\
\hline
\textbf{QPSO-LightGBM} & \textbf{93.19\(\%\)} & \textbf{93.25\(\%\)} & \textbf{93.19\(\%\)} & \textbf{93.18\(\%\)} \\
\hline
LightGBM & 89.04\(\%\) & 89.03\(\%\) & 89.03\(\%\) & 89.04\(\%\) \\
\hline
Random Forest & 74.62\(\%\) & 74.59\(\%\) & 74.59\(\%\) & 74.61\(\%\) \\
\hline
XGBoost & 67.21\(\%\) & 67.20\(\%\) & 67.20\(\%\) & 67.20\(\%\) \\
\hline
GBDT & 57.64\(\%\) & 57.62\(\%\) & 57.62\(\%\) & 57.64\(\%\) \\
\hline
SVM & 54.50\(\%\) & 54.50\(\%\) & 54.49\(\%\) & 54.50\(\%\) \\
\hline
\end{tabular}
\label{tab6}
\end{center}
\end{table}

The comparison results show that after the same processing on the same dataset, the proposed algorithm is optimal in the four indexes of accuracy, precision, recall, and F1 score.

\section{Conclusions}
The assessment of the vulnerability of cyberspace assets helps to prioritize the security of highly vulnerable network assets, which is of great significance to protect vulnerable assets, reduce the risk of asset attacks, and prevent network security losses. Based on the real network asset data, this paper uses a quantum particle swarm optimization algorithm to optimize the parameters of the LightGBM model, constructs a network asset vulnerability assessment model based on QPSO-LightGBM, and comprehensively evaluates the accuracy, precision, recall, and F1 Score of this model. The results show that this model has better performance than other models on the problem of network asset vulnerability assessment, and can realize the accurate prediction of multi-level network asset vulnerability. This paper provides a certain research basis in the field of network asset vulnerability analysis and contributes to the research on the value of network asset vulnerability.

\bibliographystyle{unsrt}
\bibliography{reference}

\end{document}